\documentclass[aps,prl,preprint,groupedaddress]{revtex4-2}

\usepackage[T1]{fontenc}
\usepackage[latin2]{inputenc}
\usepackage[letterpaper]{geometry}
\usepackage{amsmath}
\usepackage{amssymb}
\usepackage{graphicx}
\usepackage{xcolor}
\usepackage{setspace}
\usepackage{amsfonts}
\usepackage{babel}

\begin{document}

\title{Polaritons of inherently interacting molecules}

\author{G\'abor J. Hal\'asz}
\affiliation{Department of Information Technology, University of Debrecen, P.O. Box 400, H-4002 Debrecen, Hungary}

\author{Csaba F\'abri}
\affiliation{Department of Theoretical Physics, University of Debrecen, P.O. Box 400, H-4002 Debrecen, Hungary}

\author{Lorenz S. Cederbaum}
\affiliation{Theoretical Chemistry, Institute of Physical Chemistry, Heidelberg University, Im Neuenheimer Feld 229, 69120 Heidelberg, Germany}

\author{\'Agnes Vib\'ok}
\email{vibok@phys.unideb.hu}
\affiliation{Department of Theoretical Physics, University of Debrecen, P.O. Box 400, H-4002 Debrecen, Hungary}
\affiliation{ELI-ALPS, ELI-HU Non-Profit Ltd, H-6720 Szeged, Dugonics t\'er 13, Hungary}

\date{\today}

\begin{abstract}
We investigate the absorption spectra of molecular ensembles strongly coupled to a Fabry--P\'erot cavity and establish the microscopic origin of collective polaritonic bands. We show that long-range intermolecular interactions generate a large manifold of hybrid molecular--photonic states, even without disorder, vibrations, or thermal effects. The observed polariton bands are collective envelopes of these microscopic states, making photonic and molecular spectra strongly dependent on molecular arrangement. Our findings identify long-range intermolecular interactions as a key ingredient of molecular cavity QED.
\end{abstract}

\maketitle

Strong coupling between molecules and confined photonic mode of cavity
leads to the formation of mixed polaritonic states. These hybrid light-matter
states combine the characteristics of both constituents, allowing
them to become delocalized over all molecules coupled to the cavity
photon. Over the past decade, hybrid vibronic (visible) and vibrational
(infrared) polaritons have been shown to profoundly modify molecular
properties \cite{12HuScGe,16ZhChWa,16Ebbesen,15GaGaFe,17FlRuAp,17HeSp_2,18RiMaDu,18RuTaFl,18FeGaGa,18Vendrell,21GaFrCi,22FrGaFe},
including, among others, photochemical reaction rates \cite{16GaGaFe,18MuWeBa,22SuVe,24DuTiSo,25MeVe},
chemical reactivity \cite{21LiMaHu,22ScFlRo}, charge and long-range energy transfer
\cite{18DuMaRi,18ReMiGe,19ScRuAp,19ReSoGe,20MaKrHu,20XiRiDu,20LiMeSf,21CeKu,22ChDuYa}, and nonadiabatic
dynamics \cite{19PeYu,16KoBeMu,18FrGrCo,18SzHaCs_2,19CsViHa,19CsKoHa,19UlGoVe,19TrSa,20GuMu,21FaHaCe,22FaHaVi,24FaHaCe,24FaCsHa_2,25FaHaHo,25SzFaHa}.
Despite the many remarkable advances in the field, the dynamics of
the collective hybrid states formed inside cavities, as well as the
microscopic structure of the states comprising them, remain far from
being fully understood. 

In this Letter, we investigate the absorption spectrum of an ensemble
of $N$ diatomic molecules strongly coupled to a single mode
of a Fabry--P\'erot cavity. We reveal how the microscopic cavity-induced
states determine the emergence of the macroscopic collective polaritonic
state observed in the spectrum. In particular, we examine how the structure of the resulting hybrid intermolecular polaritonic state is governed primarily by long-range intermolecular interactions, and how it is further influenced by the intermolecular separation, the light-matter coupling strength, and the spatial arrangement of the molecules within the ensemble.

We investigate both the photonic and molecular absorption spectra,
corresponding to the cases in which the weak probe pulse couples either
to the cavity mode \cite{20SiPiGa} or directly to the molecular ensemble. We show
that, in an idealized system where all molecules are identically oriented
and each molecule possesses the same transition dipole moment and
therefore experiences identical light--matter coupling, most of the resulting
microscopic eigenstates exhibit a mixed molecular--photonic character.

Remarkably, these hybrid microscopic states emerge even though we
deliberately exclude all mechanisms commonly associated with the formation
of gray (or quasi-dark) states, such as thermal effects, molecular
disorder, vibrational motion, and spectral overlap \cite{24PeMeYu,22WuFiWa,25BaAsSc,26BhSoSc,22GeSe,23MaTaWe,23KhGuGe,09HaToKu,22MoYuSc,24XiXi,21MoClPe,25GoPoMo}. Instead, we demonstrate
that the mixing originates solely from the long-range intermolecular
interactions within the molecular ensemble, revealing a previously
overlooked mechanism for the formation of hybrid collective states
in strongly coupled cavity--molecule systems.

As a realistic model system, we consider an ensemble of $N$  $\mathrm{Na_{2}}$ molecules in the frozen-vibration approximation interacting with a single quantized cavity
mode. Each molecule is described as a two-level system comprising
the vibrational ground state of the electronic ground-state potential $V_\textrm{X}$ ($\mathrm{X}^{1}\Sigma_{g}^{+}; v=0$ where $v$ is the vibrational quantum number),
and the $v=6$ vibrational level of the first excited electronic
potential $V_\textrm{A}$ ($\mathrm{A}^{1}\Sigma_{u}^{+}; v=6)$.
Details of the electronic structure calculations, the parameter values,
and the numerical simulations are provided in the Supplemental Material.

The coupled cavity--molecule system can be described by the Pauli-Fierz
Hamiltonian \cite{23MaTaWe,97CoDuGr} generalized to $N$ molecules and
extended with the term of long-range intermolecular interaction:
\begin{equation}
\hat{H}_{\textrm{cm}}= 
    \sum_{i=1}^{N}\hat{H}_{\textrm{m},i}
    +\hbar\omega_{\textrm{c}}\hat{a}^{\dagger}\hat{a}
    -g(\hat{a}^{\dagger}+\hat{a})\sum_{i=1}^{N}\hat{\vec{\mu_{i}}}\vec{e_{0}}
    +\frac{g^{2}}{\hbar\omega_{\textrm{c}}}
    \bigg(\sum_{i=1}^{N}\hat{\vec{\mu_{i}}}\vec{e_{0}}\bigg)^{2}
    +\sum_{\substack{i,j=1 \\ i<j}}^{N} \hat{V}_{ij}
\label{eq:Hcm}
\end{equation}
Here, $\hat{H}_{\textrm{m},i}$ is Hamiltonian of the $i^\textrm{th}$ bare molecule, 
$\omega_{\textrm{c}}$ denotes the angular frequency of the cavity
mode, $\hat{a}^{\dagger}$ and $\hat{a}$ are photon creation and
annihilation operators, $\hat{\vec{\mu}}=\sum_{i=1}^{N}\hat{\vec{\mu_{i}}}$
corresponds to the total dipole operator of the ensemble and $\vec{e}_{0}$
is the cavity field polarization vector. The cavity--molecule coupling
is described by the coupling strength parameter $g=\sqrt{\frac{\hbar\omega_{\textrm{c}}}{2\epsilon_{0}V}}$
with $\epsilon_{0}$ and $V$ being the permittivity and quantization
volume of the cavity, respectively. The fourth and last terms of $\hat{H}_{\textrm{cm}}$
are the dipole self-energy and the long-range intermolecular-interaction
$\hat{V}_{ij}$ between the members of the ensemble, respectively.
The latter one is a dipole-dipole interaction which can be approximated
as \cite{25CeHo}:
\begin{equation}
    \hat{V}_{ij}=\frac{\hat{\vec{\mu_{i}}}\circ\hat{\vec{\mu_{j}}}}{R_{ij}^{3}}=\frac{1}{R_{ij}^{3}}\left[\hat{\vec{\mu_{i}}}\cdot\hat{\vec{\mu_{j}}}-3(\vec{u}_{ij}\cdot\hat{\vec{\mu}}_{i})(\vec{u}_{ij}\cdot\hat{\vec{\mu}}_{j})\right]
\label{eq:Coulomb}
\end{equation}
where $R_{ij}$ is the distance between the center of mass of the $i^\textrm{th}$
and $j^\textrm{th}$ molecules, and $\vec{u}_{ij}$ is a unit vector pointing
from the $i^\textrm{th}$ molecule to the $j^\textrm{th}$ molecule. 
We represent Eq. \eqref{eq:Hcm}
in the product basis of molecular eigenstates and 
Fock states $|n\rangle$ with $n=0,1$: 
$|\phi_{0}\rangle=|g_{1},\ldots,g_{N}\rangle|1\rangle$
and
$|\phi_{i}\rangle=|g_{1},\ldots,g_{i-1},e_{i},g_{i+1}
\ldots,g_{N}\rangle|0\rangle$ with $i=1,\dots,N$,
where $|g_{i}\rangle$ and $|e_{i}\rangle$ denote the ground and excited eigenstates of the $i^\textrm{th}$ molecule, respectively.
As already stated earlier, we make the choice
$|g_i\rangle = |\textrm{X},v=0\rangle$ and
$|e_i\rangle = |\textrm{A},v=6\rangle$ for each molecule.
For a homonuclear diatomic molecular ensemble with no permanent dipole moments, only transition dipole moments contribute. Consequently, the interaction term in Eq. \eqref{eq:Hcm} reduces to the resonant transition dipole-dipole interaction. The calculations are restricted
to the single-excitation manifold. The resulting Hamiltonian matrix
is given in the Supplemental Material. 

The field-dressed states $\left|\Psi_{i}^{\mathrm{FD}}\right\rangle $
are obtained by diagonalizing the Hamiltonian and can be expanded as
$\left|\Psi_{i}^{\mathrm{FD}}\right\rangle = \sum_{k=0}^{N}C_{i,k}\left|\phi_{k}\right\rangle$.
Within first-order time-dependent perturbation theory, the absorption
intensities are $I_{i}^{\mathrm{ph}}=|C_{i,0}|^{2}$ when the probe
couples to the cavity mode (photonic spectrum), and $I_{i}^{\mathrm{mol}}=\left|\sum_{k=1}^{N}C_{i,k}\right|^{2}$
when it couples to the molecular transition dipoles (molecular spectrum).
In all of the spectra shown in this work, we plot the $I_{i}^{\mathrm{ph}}$
and $I_{i}^{\mathrm{mol}}$ quantities as stick spectra together with their convolution with a Gaussian function whose  width simulates the experimental resolution,  the latter representing the experimentally  'observed' spectrum. 

\begin{figure}
\includegraphics[width=0.65\columnwidth]{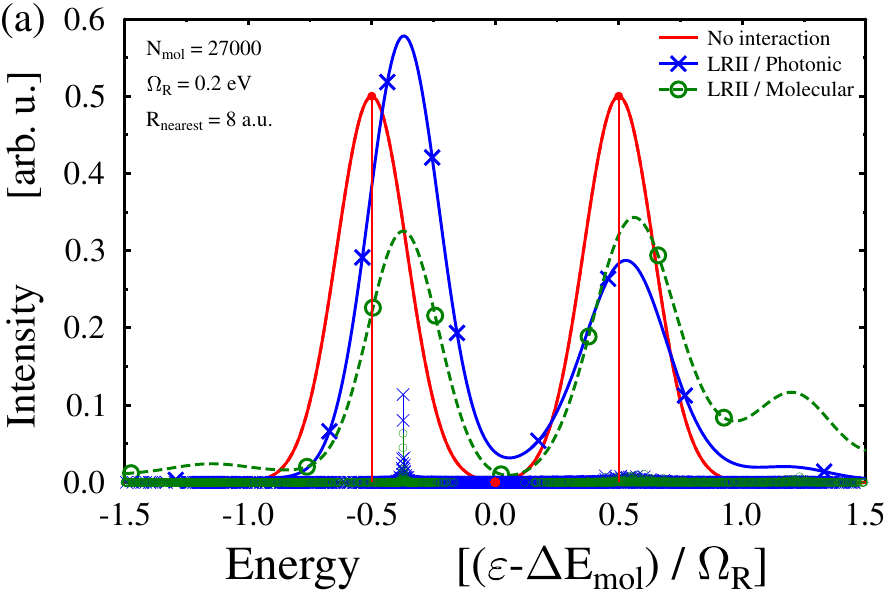}
\includegraphics[width=0.65\columnwidth]{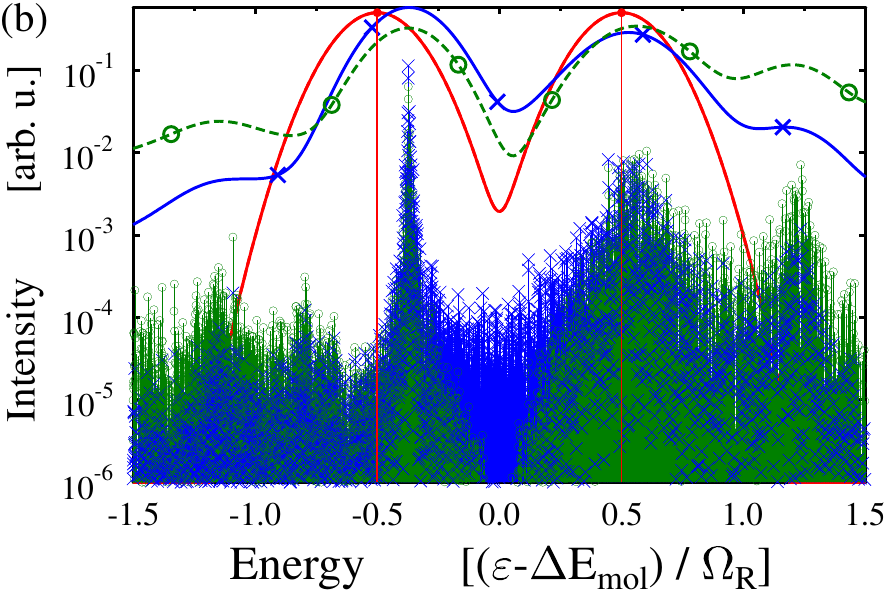}
\caption{\label{fig:1}
Cavity-dressed absorption spectra of an ensemble of $N=27000$ 
$\mathrm{Na_{2}}$ molecules arranged in a cubic close-packed (CCP)
lattice with identical orientations and an intermolecular separation
of $R = 8 ~ \textrm{a.u.}$
The cavity frequency $\omega_\textrm{c}=1.9404 ~ \textrm{eV}$ is chosen to
be resonant with the bare molecular transition energy, $\Delta E_{\mathrm{mol}}=1.9404 ~ \textrm{eV}$. The collective Rabi splitting is $\Omega_\textrm{R}=0.2 ~ \textrm{eV}$ and the
spectra are convoluted with a Gaussian of standard deviation $\sigma=\frac{1}{5\sqrt{2}}\Omega_\textrm{R}$.
The energy axis is shifted such that ($E=0$) corresponds to the bare
molecular transition energy. (a) Reference spectrum without long-range
intermolecular interactions (LRII, red solid line), together with
the photonic (blue solid line with crosses) and molecular (green dashed
line with circles) spectra including LRII. (b) To better resolve the structure of the microscopic polaritonic states, the underlying stick spectra are shown on a logarithmic scale.}

\end{figure}

It is well established that, in the strong-coupling regime, an ensemble
of $N$ two-level molecules is described by the Tavis--Cummings model \cite{68TaCu},
which predicts an absorption spectrum consisting of upper and lower
polaritons separated by $N-1$ dark molecular states. The upper and
lower polaritons are hybrid light--matter states containing both
photonic and molecular components. As a reference, the solid curve
in panel (a) of Fig. \ref{fig:1} reproduces this textbook behavior for an ensemble of $N=27000$
$\mathrm{Na}_{2}$ molecules arranged in a cubic close-packed
(CCP) lattice with {\em identical} orientations. We consider an intermolecular
separation of $R = 8 ~ \textrm{a.u.}$ and a cavity mode of $\omega_\textrm{c}=1.9404~\textrm{eV}$
which is resonant with the molecular transition. The applied collective
Rabi splitting is $\Omega_\textrm{R}=\sqrt{N}g=0.2~\textrm{eV}$. In this calculation,
the Hamiltonian includes only the first three terms of Eq. \eqref{eq:Hcm},
yielding the expected symmetric upper and lower polariton peaks. The
photonic and molecular spectra are identical in this situation. Including
the long-range intermolecular interaction (the last term in Eq. \eqref{eq:Hcm})
qualitatively modifies the absorption spectrum. Depending on whether
the probe couples to the cavity mode or to the molecular transition
dipoles, the photonic and molecular spectra (solid with cross and
dashed with circle curves in panel (a) of Fig. \ref{fig:1}, respectively) exhibit markedly
asymmetric upper and lower polariton peaks accompanied by additional
satellite features. Since all other broadening and symmetry-breaking
mechanisms---such as molecular vibrations, thermal fluctuations,
and static disorder---are deliberately excluded from the model, these
spectral features arise solely from the long-range intermolecular
interactions accounted for. To uncover their physical origin, we next
analyze the underlying discrete spectrum of microscopic eigenstates
that gives rise to the observed macroscopic absorption profiles.

The microscopic structure underlying the spectra in panel (a) of Fig. \ref{fig:1} becomes
even more evident in the logarithmic representation of panel (b) of Fig. \ref{fig:1}. Rather
than consisting of only two ideal polaritonic modes, the coupled cavity--molecule
system supports a large number of microscopic hybrid polaritonic states
generated by long-range intermolecular electrostatic interactions.
The experimentally observed collective upper and lower polariton bands
therefore emerge as the superposition of these microscopic hybrid
states. These microscopic hybrid polaritonic states are an inherent
feature of the coupled cavity--molecule system and arise solely
from long-range intermolecular electrostatic interactions.

\begin{figure}
\includegraphics[width=0.65\columnwidth]{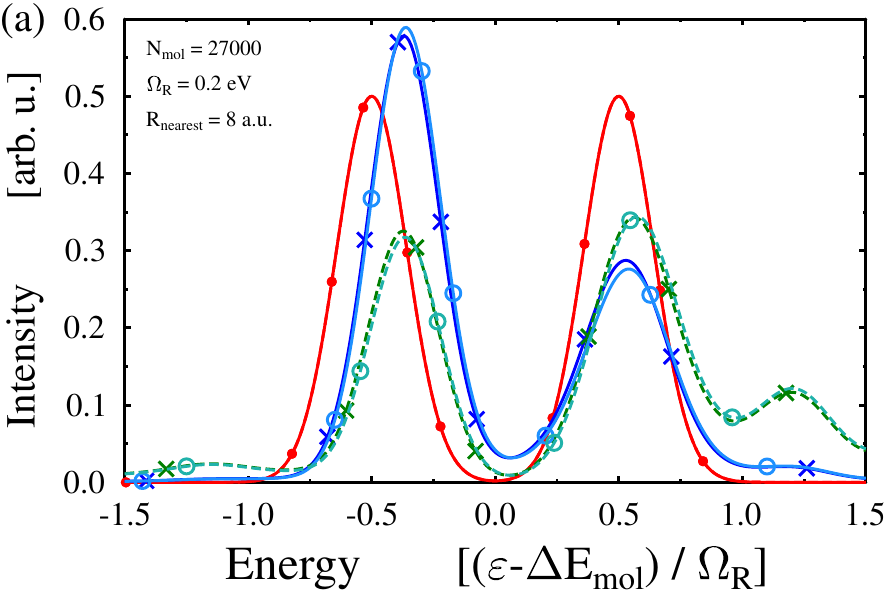}
\includegraphics[width=0.65\columnwidth]{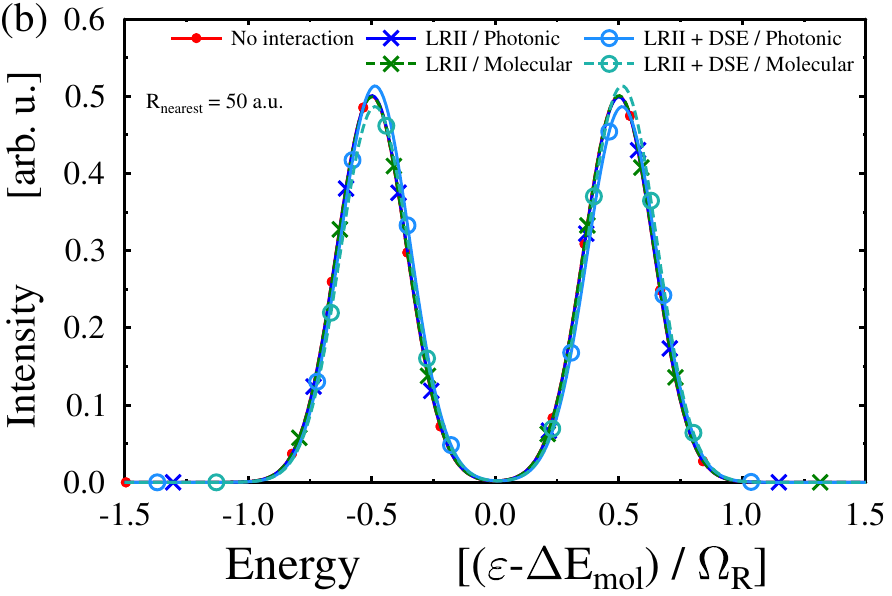}
\caption{\label{fig:2}
Cavity-dressed absorption spectra of an ensemble of $N=27000$ 
$\mathrm{Na_{2}}$ molecules arranged in a CCP lattice with identical
orientations and intermolecular separations of $R = 8 ~ \textrm{a.u.}$ (a) and $R = 50 ~ \textrm{a.u.}$
(b). All other quantities are the same as in Fig. \ref{fig:1}. Solid and dashed curves denote the photonic
and molecular spectra, respectively. In both panels, the reference spectrum without long-range intermolecular interactions (LRII, red line with dots) is compared with the spectrum including long-range intermolecular interactions (blue and green lines with crosses) and the spectrum including both long-range intermolecular interactions and the DSE term (blue and green lines with circles).
}
\end{figure}

In Fig. \ref{fig:2}, panels (a) and (b) examine the role of the dipole self-energy (DSE)
by employing the full Hamiltonian of Eq. \eqref{eq:Hcm}. We consider the same as in Fig. \ref{fig:1}, but also for $R=50 ~ \textrm{a.u.}$
Details of our highly-accurate DSE implementation are provided in the Supplemental Material.
As a reference, we also show the spectrum obtained without intermolecular
interactions, characterized by symmetric upper and lower polariton
peaks, together with the corresponding photonic and molecular spectra
including long-range intermolecular interactions.

For $R=8 ~ \textrm{a.u.}$ in panel (a) of Fig. \ref{fig:2}, the photonic and molecular spectra differ markedly
from the reference, reflecting the strong influence of long-range
intermolecular interactions. In contrast, at $R=50 ~ \textrm{a.u.}$ (see panel (b) of Fig. \ref{fig:2}), these
interactions become small and all three spectra nearly coincide.
In both cases, however, the inclusion of the DSE produces only marginal
changes. This is expected because the single-molecule light--matter
coupling $g$ is small, while the experimentally relevant collective
Rabi splitting is kept fixed at $\Omega_\textrm{R}=0.2 ~ \textrm{eV}$ throughout.

\begin{figure}
\includegraphics[width=0.65\columnwidth]{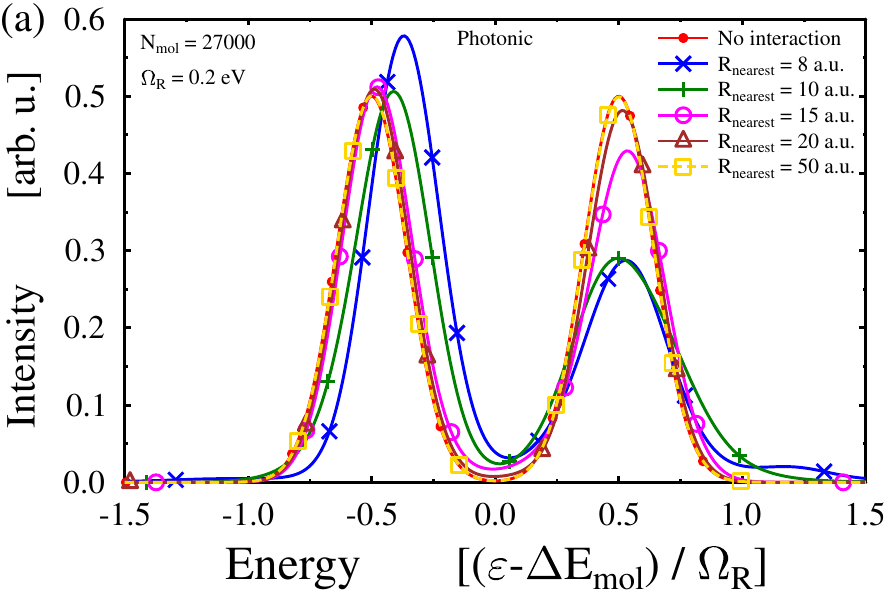}
\includegraphics[width=0.65\columnwidth]{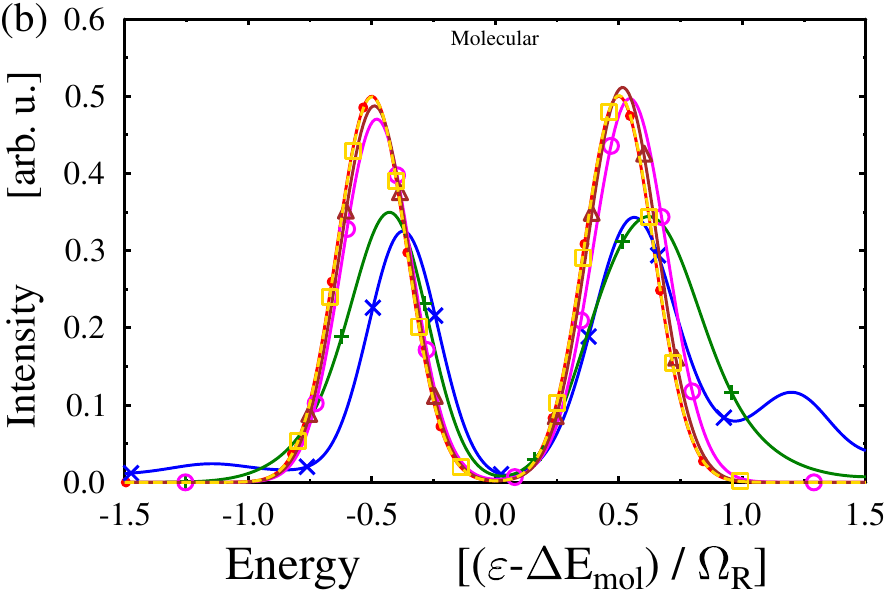}
\caption{\label{fig:3}
Cavity-dressed absorption spectra of an ensemble of $N=27000$ 
$\mathrm{Na_{2}}$ molecules arranged in a CCP lattice with identical
orientations and intermolecular separations of $R=8,10,15,20$ and $50
~\textrm{a.u.}$
All other quantities are the same as in Fig. \ref{fig:1}.
Panels (a) and (b) show the photonic
and molecular absorption spectra, respectively, including long-range
intermolecular interactions (LRII). In both panels, the reference
spectrum obtained without LRII is also shown.}
\end{figure}

Panels (a) and (b) of Fig. \ref{fig:3} show the effect of intermolecular separation on
the absorption spectra for $R=8,10,15,20$ and $50 ~ \textrm{a.u.}$, using the same
model and parameters as above. For clarity, the photonic and molecular
spectra are presented separately in panels (a) and (b) of Fig. \ref{fig:3}, respectively.
Both spectra exhibit pronounced modifications at $R=8$ and $10 ~ \textrm{a.u.}$,
demonstrating the strong influence of long-range intermolecular interactions.
The effect remains discernible at $R=15 ~ \textrm{a.u.}$ but becomes small
for $R\ge20 ~ \textrm{a.u.}$, where the spectra are nearly indistinguishable
from those obtained without intermolecular interactions.

\begin{figure}
\includegraphics[width=0.65\columnwidth]{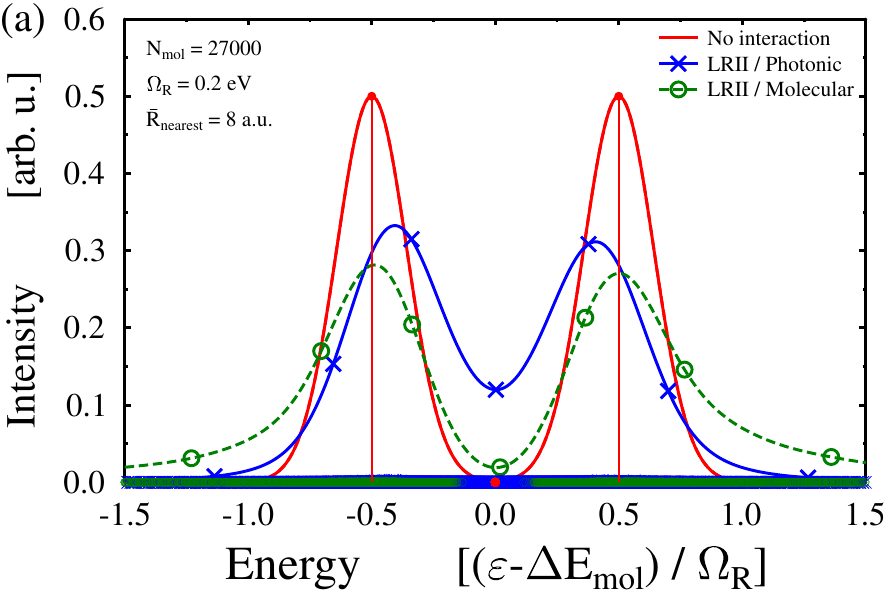}
\includegraphics[width=0.65\columnwidth]{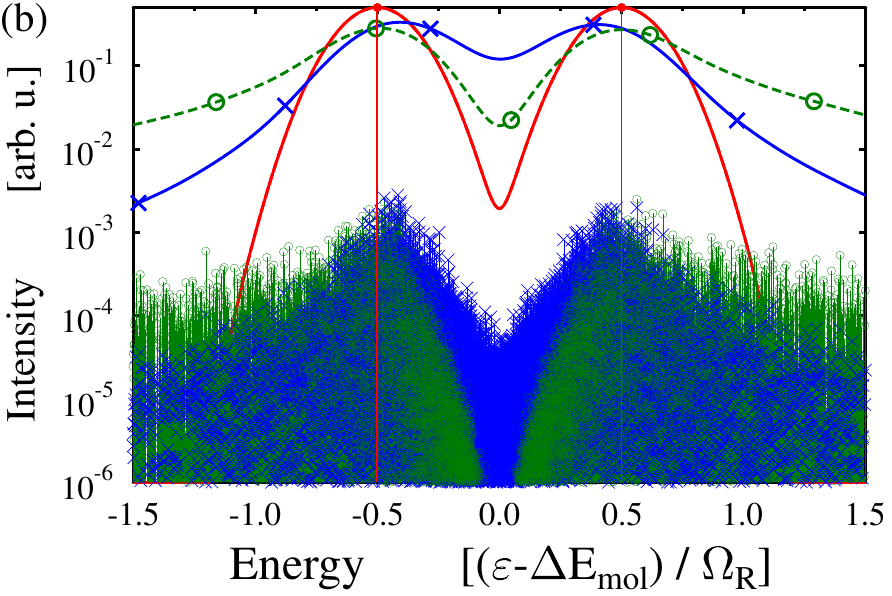}
\caption{\label{fig:4}
Cavity-dressed absorption spectra for a randomly arranged molecular
ensemble of $N=27000$ $\mathrm{Na_{2}}$ molecules. All
other quantities are the same as in Fig. \ref{fig:1}.}
\end{figure}

Panels (a) and (b) of Fig. \ref{fig:4} present the results for a randomly arranged molecular
ensemble. The molecules are uniformly distributed within a cubic volume
whose size is chosen to reproduce the prescribed mean nearest-neighbor
distance. As in the CCP lattice in panel (a) of Fig. \ref{fig:1}, the photonic and molecular
upper and lower polariton bands substantially differ from each other and from the reference spectrum.
However, in the random ensemble these deviations occur in the same
direction, making the photonic and molecular spectra much more similar
to each other than in the ordered CCP case. Moreover, the pronounced
satellite features observed in panel (a) of Fig. \ref{fig:1} are largely absent and substituted by long tails. 
An important consequence is that only long-range intermolecular interactions make the cavity spectra sensitive to the spatial arrangement of the molecules. In their absence, the spectra are independent of molecular arrangement.

The logarithmic representation in panel (b) of Fig. \ref{fig:4} reveals the microscopic
polaritonic structure even more clearly than in the ordered lattice.
The microscopic hybrid states exhibit a substantially smaller photonic
component---typically one to two orders of magnitude weaker than
in the CCP configuration. Consequently, they are expected to couple
much less efficiently to cavity losses, giving rise to longer-lived
polaritonic states and making them particularly attractive for spectroscopic
studies.

In conclusion, we have shown that the macroscopic polaritonic bands observed in cavity absorption spectra are collective envelopes arising from an extremely large number of microscopic hybrid polaritonic states. These states originate from long-range intermolecular interactions and therefore constitute an intrinsic property of molecular ensembles under strong light--matter coupling. In their absence, the photonic and molecular spectra are identical and independent of the molecular arrangement, whereas intermolecular interactions make the spectra strongly arrangement dependent and can lead to pronounced differences between photonic and molecular absorption. 
Since most cavity QED experiments are performed in the condensed phase, where such interactions are unavoidable, they should be included in realistic theoretical descriptions. Finally, the microscopic polaritonic states exhibit only a small photonic admixture, suggesting a weaker coupling to cavity losses, longer lifetimes, and enhanced prospects for their direct spectroscopic observation. These findings provide a microscopic interpretation of collective polaritonic spectra and establish long-range intermolecular interactions as a key ingredient of molecular cavity QED.

\begin{acknowledgments}
The authors are indebted to NKFIH for funding (Grant No. K146096).
The ELI ALPS project (GINOP-2.3.6-15-2015-00001) is supported by the
European Union and co-financed by the European Regional Development
Fund. 
This paper was supported by the J\'anos Bolyai Research Scholarship of the Hungarian Academy of Sciences as well
as by the University of Debrecen Program for Scientific Publication.
\end{acknowledgments}

\bibliography{Coulomb_Inherent}

\clearpage

\begin{center}
\textbf{\Large{}Supplemental Material}{\Large\par}
\par\end{center}

\part*{1. Electronic structure details of the Na$_2$ molecule}

Below we provide the electronic structure details of the Na$_2$ molecule, considered in the main text.
\vspace{0.2cm}

\begin{figure}[ht]
\begin{center}
\includegraphics[clip,width=8cm]{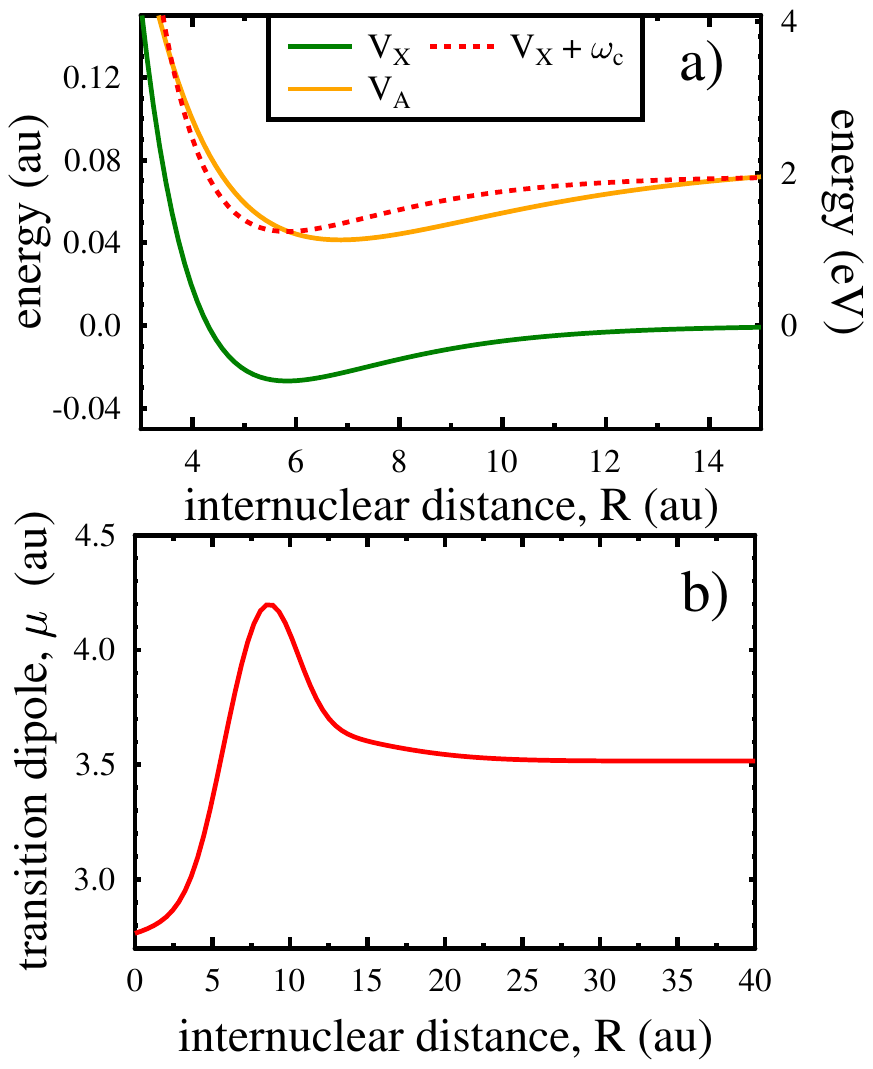} 
\caption{
(a) Potential energy curves of the ground (X) and first excited (A) electronic states of the Na$_2$ molecule, considered in the main text. 
The broken line represents the ground state potential curve shifted by the cavity photon energy $\omega_c = 1.9404$ eV. 
(b) Transition dipole moment between the X and A electronic states.}
\end{center}
\label{F1}
\end{figure}

The potential curves are taken from Ref.~\cite{93MaMiDu}:
\begin{equation}\label{eq:potx}
V_\textrm{X} (R) = D_\textrm{X} [ e^{-\alpha_\textrm{X} (R-R_\textrm{X})} - 1 ]^2 - D_\textrm{X} 
\end{equation}
\begin{equation}\label{eq:pota}
V_\textrm{A} (R) = D_\textrm{A} [ e^{-\alpha_\textrm{A} (R-R_\textrm{A})} - 1 ]^2 - D_\textrm{A} + V_\textrm{sh}
\end{equation}
where $R$ is the internuclear separation and the parameter values are given as:

$D_\textrm{X} = 5892$ cm$^{-1}$

$R_\textrm{X} = 5.83$ a.u.

$O_\textrm{X} = 159.1$ cm$^{-1}$

$\alpha_\textrm{X} = 0.5 O_\textrm{X} \sqrt{2 M / D_\textrm{X}}$

$D_\textrm{A} = 8284$ cm$^{-1}$

$R_\textrm{A} = 6.86$ a.u.

$O_\textrm{A} = 117.5$ cm$^{-1}$

$\alpha_\textrm{A} = 0.5 O_\textrm{A} \sqrt{2 M / D_\textrm{A}}$,

in addition, the reduced mass is $M=20953.89282$ a.u. $=11.49488464$ amu and $V_\textrm{sh}=0.07917$ a.u. is the vertical energy shift.

With these potentials, the energies of the low-lying vibrational states are the following ($v$ denotes the vibrational quantum number): \\
  {  \centering
    \begin{tabular}{c|cccccccccc}
      $v$             & 0         & 1         & 2         & 3         & 4         & 5         & 6        \\
\hline 
       $E_\textrm{X}(v)$ / eV  & -0.720685 & -0.701226 & -0.682032 & -0.663105 & -0.644445 & -0.626051 & -0.607923 \\
       $E_\textrm{A}(v)$ / eV  & ~1.134511 & ~1.148976 & ~1.163338 & ~1.177596 & ~1.191751 & ~1.205802 & ~1.219751 
    \end{tabular}
}

The transition dipole moment values (aligned with the molecular axis) were taken from Ref.~\cite{81ZeVeVu} and fitted with the formula: 
\begin{equation}\label{eq:TDM}
\mu (R) = \mu_i +(\mu_f - \mu_i) \frac{1}{2} [1 + \mathrm{tanh}(a (R-b))] + c \cdot e^{-d (R-f)^2} + g \cdot e^{-h (R-k)^2} 
\end{equation}

using the following parameter values:

$\mu_i = 2.655$ a.u.

$\mu_f = 3.5165$ a.u.

$a=0.4758$ a.u.

$b=4.97237$ a.u.

$c=0.545563$ a.u.

$d=0.136289$ a.u.

$f=8.54772$ a.u.

$g=0.168368$ a.u.

$h=0.0102951$ a.u.

$k=6.87108$ a.u.

\part*{2. Details of the numerical simulations}

It was shown in Ref.~\cite{25CsSzVi} for the case of weak coupling that the only vibrational states which are 
highly populated are the $v=0$ vibrational state in the ground (X) electronic state and $v=6$ vibrational state in the first excited (A) 
electronic state. 
Based on this finding we restrict ourselves to include only these two 
eigenstates into our current simplified model for the ensemble of 
$\mathrm{Na_{2}}$ molecules. The energy difference and the effective transition dipole moment between them 
are $\Delta E = E_\textrm{A}(v=6) - E_\textrm{X}(v=0) = 1.9404$ eV and
$ \langle \textrm{X},v=0|\hat\mu|\textrm{A},v=6\rangle = 1.21955429 $ a.u., respectively,
where $|\textrm{X},v=0\rangle$ and $|\textrm{A},v=6\rangle$ denote the ground (X) and excited (A) 
electronic state wave functions of the molecule in the given vibrational state.

We consider an ensemble of $N$ $\mathrm{Na_{2}}$ molecules using the simplification introduced in the previous paragraph.
Thus, we assume that molecular eigenstates
$|\textrm{X},v=0\rangle$ and $|\textrm{A},v=6\rangle$
interact with the quantized electromagnetic mode of a cavity (with photon frequency $\omega_{c}$). 
The coupled cavity--molecule system can be described by the Pauli-Fierz
Hamiltonian \cite{23MaTaWe,97CoDuGr} generalized to $N$ molecules and
extended with the term describing long-range intermolecular interactions:
\begin{equation}
\hat{H}_{\textrm{cm}}= 
    \sum_{i=1}^{N}\hat{H}_{\textrm{m},i}
    +\hbar\omega_{\textrm{c}}\hat{a}^{\dagger}\hat{a}
    -g(\hat{a}^{\dagger}+\hat{a})\sum_{i=1}^{N}\hat{\vec{\mu_{i}}}\vec{e_{0}}
    +\frac{g^{2}}{\hbar\omega_{\textrm{c}}}
    \bigg(\sum_{i=1}^{N}\hat{\vec{\mu_{i}}}\vec{e_{0}}\bigg)^{2}
    +\sum_{\substack{i,j=1 \\ i<j}}^{N} \hat{V}_{ij}
\label{eq:Hcm_sm}
\end{equation}
where $\hat{H}_{\textrm{m},i}$ is Hamiltonian of the $i^\textrm{th}$ bare molecule,
$\omega_{\textrm{c}}$
denotes the angular frequency of the cavity mode, $\hat{a}^{\dagger}$
and $\hat{a}$ are photon creation and annihilation operators, $\hat{\vec{\mu}}=\sum_{i=1}^{N}\hat{\vec{\mu_{i}}}$
corresponds to the total dipole operator of the ensemble, and $\vec{e}_{0}$
is the cavity field polarization vector. The cavity--molecule coupling
is described by the coupling strength parameter $g=\sqrt{\frac{\hbar\omega_{\textrm{c}}}{2\epsilon_{0}V}}$
with $\epsilon_{0}$ and $V$ being the permittivity and quantization
volume of the cavity, respectively. The fourth and last terms of $\hat{H}_{\textrm{cm}}$
are the dipole self-energy and the long-range intermolecular-interaction
$\hat{V}_{\textrm{ij}}$ between members of the ensemble, respectively.
The latter one is a dipole-dipole interaction which can be approximated as\cite{25CeHo}
\begin{equation}
    \hat{V}_{ij}=\frac{\hat{\vec{\mu_{i}}}\circ\hat{\vec{\mu_{j}}}}{R_{ij}^{3}}=\frac{1}{R_{ij}^{3}}\left[\hat{\vec{\mu_{i}}}\cdot\hat{\vec{\mu_{j}}}-3(\vec{u}_{ij}\cdot\hat{\vec{\mu}}_{i})(\vec{u}_{ij}\cdot\hat{\vec{\mu}}_{j})\right]
\label{eq:Coulomb_sm}
\end{equation}
where $R_{ij}$ is the distance between the center of mass of the $i^\textrm{th}$
and $j^\textrm{th}$ molecules, and $\vec{u}_{ij}$ is a unit vector pointing
from the $i^\textrm{th}$ molecule to the $j^\textrm{th}$ molecule. 

For an ensemble of $N$ molecules, the Hamiltonian of Eq. \eqref{eq:Hcm_sm} can be represented in a basis constructed from field-free molecular product states $(\left|g_{1},g_{2},...,g_{N}\right\rangle ,\left|e_{1},g_{2},...,g_{N}\right\rangle, \dots, \left|g_{1},g_{2},...,e_{N}\right\rangle )$
and Fock states $\left|n\right\rangle$ of the radiation
field with $n=0,1$, that is,
\begin{align}
\left|\phi_{0}\right\rangle  & =\left|g_{1},g_{2},...,g_{N}\right\rangle \left|1\right\rangle \nonumber \\
\left|\phi_{1}\right\rangle  & =\left|e_{1},g_{2},...,g_{N}\right\rangle \left|0\right\rangle \nonumber \\
\left|\phi_{2}\right\rangle  & =\left|g_{1},e_{2},...,g_{N}\right\rangle \left|0\right\rangle \label{eq:bases}\\
 & \vdots\nonumber \\
\left|\phi_{N}\right\rangle  & =\left|g_{1},g_{2},...,e_{N}\right\rangle \left|0\right\rangle \nonumber 
\end{align}
where $|g_{i}\rangle$ and $|e_{i}\rangle$ denote the ground and excited 
eigenstates of the $i^\textrm{th}$ molecule, respectively. 
In the current work, we apply
$|g_i\rangle = |\textrm{X},v=0\rangle$ and
$|e_i\rangle = |\textrm{A},v=6\rangle$
for each molecule.
The calculations are restricted to the single-excitation manifold and since the homonuclear diatomic $\textrm{Na}_2$ molecule is considered, the permanent dipole moments vanish in both electronic states. 
The resulting Hamiltonian matrix is given as 
\begin{gather}
\hat{H}_{\textrm{cm}}  \ensuremath{=
\begin{bmatrix}NE_\textrm{g}+\hbar\omega_{c} & -g \mu_{1} & -g \mu_{2} & ... & ... & -g \mu_{N}\\
-g \mu_{1} & (N-1)E_\textrm{g}+E_\textrm{e} & W_{12} & ... & ... & W_{1N}\\
-g \mu_{2} & W_{21} & (N-1)E_\textrm{g}+E_\textrm{e} & ... & ... & W_{2N}\\
... & ... & ... & ... & ... & ...\\
... & ... & ... & ... & ... & ...\\
-g \mu_{N} & W_{N1} & W_{N2} & ... & ... & (N-1)E_\textrm{g}+E_\textrm{e}
\end{bmatrix}} \nonumber \\
+ \frac{g^2}{\hbar \omega_\textrm{c}} \begin{bmatrix} \sum_{i=1}^N D_{\textrm{g},i} & 0 & 0 & ... & ... & 0\\
0 & \sum_{i=2}^N D_{\textrm{g},i}+D_{\textrm{e},1} & 2\mu_1\mu_2 & ... & ... & 2\mu_1\mu_N\\
0 & 2\mu_2\mu_1 & \sum_{i=1,i\ne2}^N D_{\textrm{g},i}+D_{\textrm{e},2} & ... & ... & 2\mu_2\mu_N \\
... & ... & ... & ... & ... & ...\\
... & ... & ... & ... & ... & ...\\
0 & 2\mu_N\mu_1 & 2\mu_N\mu_2 & ... & ... & \sum_{i=1}^{N-1}D_{\textrm{g},i}+D_{\textrm{e},N}
\end{bmatrix} \label{eq:Hcm_matrix} 
\end{gather}
where in the first matrix, $E_\textrm{g}$ and $E_\textrm{e}$ are the ground and excited molecular energies, corresponding to molecular eigenstates $|g_i\rangle$ and $|e_i\rangle$, respectively.
In addition, the transition dipole moment is denoted by 
$\mu_i = \langle g_i | \hat{\vec{\mu_i}} \vec{e_0} | e_i \rangle$,
while $W_{ij}$ is the matrix element of the dipole-dipole interaction between two molecules:
\begin{equation}
W_{ij}=\frac{1}{R_{i,j}^{3}}[\vec{\mu}_{i}\cdot\vec{\mu}_{j}-3(\vec{u}_{ij}\cdot\vec{\mu}_{i})(\vec{u}_{ij}\cdot\vec{\mu}_{j})]
\label{eq:Coulomb_energy}
\end{equation}
where $\vec{\mu}_i = \langle g_i | \hat{\vec{\mu_i}} | e_i \rangle$ is the transition dipole moment vector.
In the second matrix in Eq. \eqref{eq:Hcm_matrix}, the DSE terms are specified with
$D_{\textrm{g},i}= \langle g_i | (\hat{\vec{\mu_i}} \vec{e_0})^2 | g_i \rangle$
and
$D_{\textrm{e},i}= \langle e_i | (\hat{\vec{\mu_i}} \vec{e_0})^2 | e_i \rangle$. 
For simplicity, we assume an
ensemble of $N$ identical molecules with fixed, identical orientations,
such that all molecular axes are parallel to each other and aligned
with the polarization direction of the cavity electric field.

Following earlier work reported in Ref.~\cite{25FaHaHo}, we have evaluated electronic matrix elements
$\langle \textrm{X} | \hat{\mu}_k \hat{\mu}_l | \textrm{X} \rangle$
and
$\langle \textrm{A} | \hat{\mu}_k \hat{\mu}_l | \textrm{A} \rangle$
($k,l=x,y,z$), required by the DSE term,
as a function of the internuclear separation for the $\textrm{Na}_2$ molecule.
Due to symmetry, DSE matrix elements between the ground (X) and excited (A) electronic states are identically zero, that is,
$\langle \textrm{X} | \hat{\mu}_k \hat{\mu}_l | \textrm{A} \rangle = 0$.
In addition, if the molecular axis coincides with the body-fixed $z$ axis,
$\langle \alpha | \hat{\mu}_x^2 | \alpha \rangle = \langle \alpha | \hat{\mu}_y^2 | \alpha \rangle \ne \langle \alpha | \hat{\mu}_z^2 | \alpha \rangle$ 
($\alpha = \textrm{X},\textrm{A}$) and
$\langle \alpha | \hat{\mu}_k \hat{\mu}_l | \alpha \rangle = 0$
for $k \ne l$.
For the X state, matrix elements 
$\langle \textrm{X} | \hat{\mu}_k \hat{\mu}_l | \textrm{X} \rangle$
were computed at the CCSD/cc-pVTZ level of theory using one-electron and two-electron reduced density matrices (see Ref.~\cite{25FaHaHo} for more information).
For the A state, the EOM-CCSD/cc-pVTZ level of theory was employed and matrix elements 
$\langle \textrm{A} | \hat{\mu}_k \hat{\mu}_l | \textrm{A} \rangle$
were computed by numerical differentiation of the electronic energy of an appropriately perturbed Hamiltonian.
In both cases, the PySCF program package \cite{18SuBeBl,20SuZhBa} was used. We note that DSE matrix elements were obtained directly and no approximations were made in addition to the standard approximations commonly used in quantum chemistry.

Let us now calculate the $\left|\Psi_i^\mathrm{FD}\right\rangle $
field-dressed states of the coupled cavity--molecule system. These
states are the eigenstates of the Hamiltonian matrix of Eq. \eqref{eq:Hcm_matrix}.
The field-dressed states can be written as the linear combination
of products of field-free molecular eigenstates and Fock states
of the cavity mode, that is,
\begin{equation}
\left|\Psi_{i}^\mathrm{FD}\right\rangle = \sum_{k=0}^{N}C_{i,k}\left|\phi_{k}\right\rangle
\label{eq:Field-dressed state}
\end{equation}
with $i=0,1,2,...,N$.
One can calculate the absorption spectrum of the coupled system with
respect to a weak probe pulse by applying first-order time-dependent
perturbation theory. Then one can calculate the transition
amplitudes between two field-dressed states. 
In this work, the initial state of transitions to the single-excitation manifold is chosen as
$|\Psi_0 \rangle = \left|g_{1},g_{2},...,g_{N}\right\rangle \left|0\right\rangle$.
The absorption spectrum can be calculated in two different ways:
(i) if the cavity photon couples
directly to the electromagnetic field of the weak probe laser pulse,
the resulting spectrum is referred to as the photonic spectrum with intensities 
\begin{equation}
   	I_{i}^{\mathrm{ph}}=|C_{i,0}|^{2},~~~i=0,...,N.
\label{eq:trans_amp_photonic}
\end{equation}
Alternatively, (ii) if the molecular transition dipole moment couples
to the external probe field, the resulting spectrum is the molecular
spectrum with
\begin{equation}
	I_{i}^{\mathrm{mol}}=\left|\sum_{k=1}^{N}C_{i,k}\right|^{2},~~~i=0,...,N.
\label{eq:trans_amp_molecular}
\end{equation}

\end{document}